\documentclass[useAMS,usenatbib]{mn2e} 

\usepackage{graphicx}
\usepackage{color} 
\usepackage{longtable} 
\usepackage{times}

\title[The M$_{BH}$- Doppler boosted emission relation]{The connection
  between black hole mass and Doppler boosted emission in BL Lacertae
  type objects.}

 \author[J. Le\'on-Tavares et al.]{J. Le\'on-Tavares$^{1}$\thanks{E-mail:
leon@kurp.hut.fi}, E. Valtaoja$^{2}$, V. H. Chavushyan$^{3}$, M. Tornikoski$^{1}$, C. A\~norve$^{3}$, 
 \newauthor
  E. Nieppola $^{1, 4}$, A.  L\"ahteenm\"aki$^{1}$\\ \\
 $^{1}$ Aalto University Mets\"ahovi Radio Observatory,  Mets\"ahovintie 114, FIN-02540
   Kylm\"al\"a, Finland.\\
 $^{2}$  Tuorla Observatory, Department  of Physics and Astronomy, University of Turku, 20100 Turku, Finland.\\
 $^{3}$ Instituto Nacional de Astrof\'{\i}sica \'Optica y
    Electr\'onica, Apartado Postal 51 y 216, 72000 Puebla, M\'exico\\
 $^{4}$ Finnish Center for Astronomy with ESO (FINCA)), University of Turku,
  V\"ais\"al\"antie 20, FI-Piikki\"o, Finland 
 }
\begin{document}

\maketitle
\begin{abstract}

  We investigate the relationship between black hole mass
  (M$_{BH}$) and Doppler boosted emission for BL Lacertae type objects
  (BL Lacs) detected in the SDSS and FIRST surveys. The synthesis of
  stellar population and bidimensional decomposition methods allows us
  to disentangle the components of the host galaxy from that of the
  nuclear black hole in their optical spectra and images,
  respectively. We derive estimates of black hole masses via stellar
  velocity dispersion and bulge luminosity. We find that masses
  delivered by both methods are consistent within errors. There is no
  difference between the black hole mass ranges for high-synchrotron
  peaked BL Lacs (HBL) and low-synchrotron peaked BL Lacs (LBL). A
  correlation between the black-hole mass and radio, optical and X-ray
  luminosity has been found at a high significance level. The
  optical-continuum emission correlates with the jet luminosity as
  well. Besides, X-ray and radio emission are correlated when HBLs and
  LBLs are considered separately. Results presented in this work: (i)
  show that the black hole mass does not decide the SED shapes of BL
  Lacs, (ii) confirm that X-ray and optical emission is associated to
  the relativistic jet, and (iii) present evidence of a relation
  between M$_{BH}$ and Doppler boosted emission, which among BL Lacs
  may be understood as a close relation between faster jets and more
  massive black holes.

\end{abstract}


\section{INTRODUCTION}

BL Lac type objects and flat-spectrum radio quasars populate the class
of Active Galactic Nuclei (AGN) denominated as blazars.  The spectral
energy distribution (SED) in BL~Lacs usually shows a doble-peaked
shape. Depending on whether the first peak (synchrotron contribution)
lies in the optical-IR bands or if the peak is located at X-ray
regime, the BL~Lac is classified as low-peaked BLLac (LBL) or
high-peaked BLLac (HBL), respectively.  The large majority of AGN
detected by the \textit{Fermi Large Area Telescope} in its first 11
months of sky survey are blazars and about $40\%$ of these sources are
classified as BL Lac type objects (Abdo et al.  2010).  As has been
shown by recent studies (Lister et al. 2009, Savolainen et al. 2010,
Tornikoski et al. 2010), $\gamma$-ray blazars tend to have
preferentially higher Doppler factors. Using a sample of radio-loud
AGN (including mostly blazars), Arshakian et al. (2005) and Valtaoja
et al. (2008) have drawn attention to a positive correlation between
black hole mass and Doppler factors.  According to these two pieces of
observational evidence, we may rise the following question: should we
expect a population of heavier black holes for those blazars detected
at high-energies? In order to address this question, it is extremely
important to obtain reliable estimates of the black hole masses in
blazars.

Besides, in order to successfully interpret and model the variability
in blazars, it is not sufficient to reproduce the observed spectral
energy distributions (SEDs), the variability must occur on a physical
timescales that is consistent with the chosen model. Thus, in order to
discriminate between theoretical models and get a true estimation of
the physical scales in the central engine of blazars, a reliable
estimation of the black-hole mass must be known.

When measuring black hole masses in strongly beamed radio-loud sources
(i.e. blazars) using the gas in the broad line region (BLR) and
assuming its virial motion (Kaspi 2000, Peterson et al. 2004), we
should be aware of the potential biases in the measurements: (i) a
possible flat-geometry of the BLR would introduce an orientational
dependence of the FWHM, (ii) the contamination of the optical
continuum emission by non-thermal radiation from the jets and (iii)
non-virialized motions in the BLR (Arshakian et al. 2010,
Le\'on-Tavares et al. 2010, Ili\'c et al. 2009). All these potential
biases make black hole estimations in radio-loud AGN a considerable
challenge.

However, nature is particularly kind and provides us the BL Lac type
objects, which are those blazars where the host galaxy is not outshone
by the AGN emission. Thus, enabling us to study radio-loud AGN
host-galaxy properties and obtain reliable and bias-free estimations
of the black hole mass via the tight relation between the mass of
massive black holes and the stellar velocity dispersion (Ferrarese et
al 2001; Tremaine et al.  2002). 

In this work we perform measurements of the velocity dispersion and
bulge magnitudes in a sample of BL Lacs to enable reliable estimations
of black hole masses. By comparing our black hole estimates to
multiwavelegth luminosities and broad-band spectral indices, we
re-examine the relationship between black hole mass and Doppler
factors.  In the following, we use a $\Lambda$CDM cosmology with values within $1\sigma$ of the WMAP results (Komatsu et al. 2009); in particular,   H$_{0}$=71 km s$^{-1}$ Mpc$^{-1}$, $\Omega_{m}=0.27$, $\Omega_{\Lambda}=0.73$.

\section{THE SAMPLE}

Our BL Lac sample consists of the sub-sample of objects in the large
sample of BL Lacs from SDSS and FIRST (Plotkin et al. 2008) for which
reliable host galaxy and AGN decomposition has been performed in the
SDSS spectrum and optical images. The Plotkin et al. (2008) sample of
BL Lacs was selected jointly from SDSS DR5 optical spectroscopy and
the FIRST radio imaging. Their selection criteria was based on
positional matching between SDSS and FIRST surveys, and spectral
constraints commonly used in recent works to classify and characterize
BL Lacs, such as no exhibition of broad emission lines and a Ca II H/K
break measured depression $ C~\leq~0.4$.

Within the 256 higher confidence BL Lac candidates in the original
sample, we have selected 179 objects in the redshift interval z$\sim [
0.06, 0.50]$, within which strong stellar absorption lines can be
measured reliably. As such, it will be in the following identified as
the original BL Lac sample. In order to separate the spectral
contribution of the host galaxy and AGN, we used the spectral
synthesis of the stellar component method. This allows the star
formation history, stellar velocity dispersion of the host galaxy and
continuum spectra of AGN to be modeled and recovered (see section 3.1
for details). After a visual inspection of the modeled spectra, 101
out of 179 objects have been excluded from the analysis because of the
poor fitting in the spectral range available, leaving us with 78 BL
Lacs. We found that the spectra fitting goodness depends mainly on the
signal-to-noise ratio and the rather moderate contribution of AGN
optical continuum emission ($< 80\%$). Based on the X-ray to radio
flux ratio (Padovani and Giommi, 1995), we may classify our sample in
45 HBL and 33 LBL.

\begin{figure}
  \includegraphics[width=\columnwidth]{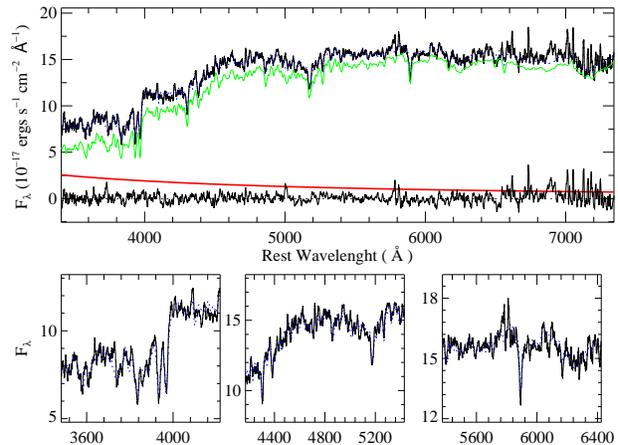}
  \caption{\emph{Top panel} Spectral synthesis for the BL Lac type
    object J000157.23-103117.3, performed by STARLIGHT (blue line)
    using 150 SSP and six power laws to simulate the AGN continuum
    emission. The observed spectra (top black line), host galaxy model
    (green line), AGN component (red line) and residuals (bottom
    black line) are shown. \emph{Bottom panel: } We show details of
    the fittings around the absorption features.}
\end{figure}

\section{MEASUREMENTS}

\subsection{Stellar velocity dispersion}

The 179 BL Lac spectra from the original sample have been retrieved
from SDSS-DR7 and corrected for Galactic extinction using the maps of
Schlegel, Finkenber \& Davis (1998). Then they are brought to the rest
frame and resampled from 3400 to 9100 \AA\ in steps of 1\AA\ with a
flux normalization by the median flux in the 4010-4060 \AA\  region. We
use the stellar population synthesis code
\textmd{STARLIGHT}\footnote{http://www.starlight.ufsc.br/} to obtain
the best fit to an observed spectrum $O_{\lambda}$ , taking into
account the corresponding error $\sigma_{obs}$. The best fit is a
combination of single stellar populations (SSP) from the evolutionary
synthesis models of Bruzual \& Charlot (2003) and power-laws to
represent the AGN continuum emission.

The code finds the minimum $\chi^{2}$,
\begin{equation}
 \chi^{2} = \sum_{\lambda} \left(  \frac{O_{\lambda}- M_{\lambda}} {\sigma_{obs}}\right)
\end{equation}
where $M_{\lambda}$ is the model spectrum (SSP and power-laws),
obtaining the corresponding physical parameters of the modeled
spectrum: Star formation history, $x_{j}$, as a function of a base
of SSP models normalized at $\lambda_{0}$, $b_{j,\lambda}$,
extinction coefficient of predefined extinction laws,
$r_{\lambda}$, and velocity dispersion $\sigma_{\star}$ which obeys
the relation:

  \begin{equation}
   M_{\lambda}= M_{\lambda 0} \left ( \sum_{j=1}^{N_{SSP}} x_{j}, b_{j,
   \lambda} r_{\lambda} \right) \otimes G(v_{\star}, \sigma_{\star})
  \end{equation}

In order to model the line-of-sight stellar motions, the code uses a
Gaussian distribution of $G$ centered at the velocity $v_{\star}$ with
dispersion $\sigma_{\star} $. We use a base of 150 SSPs plus 6 power
laws in the form F($\lambda$) = 10$^{20}$($\lambda$ / 4020)$^\beta$,
where $\beta$= -0.5, -1, -1.5, -2, -2.5, -3.  Each SSP spans six
metallicities, Z = 0.005, 0.02, 0.2, 0.4, 1 and 2.5, $Z_{\odot}$, with
25 different ages between 1~Myr and 18~Gyr. Extinction in the galaxy
is taken into account in the synthesis, assuming that it arises from a
foreground screen with the extinction law of Cardelli et al (1989).

A detailed description of the \textmd{STARLIGHT} code can be found in
the publications of the SEAGal collaboration (Cid Fernandes et
al. 2005; Mateus et al. 2006; Cid-Fernandes et al. 2007; Asari et
al. 2007). We fit all the wavelength range available in the observed
spectra (3800-9100 \AA ) . To estimate the reliable starlight
contribution to the optical spectrum we used a uniform weighting to
fit all the absorption features within the spectral range mentioned
above.  When fitting the observed spectrum, we combined a set of
power-laws (to account for the AGN continuum luminosity) with stellar
populations. This allows the reliable simultaneous decomposition of
the host galaxy and AGN optical continuum emission to be
performed. However, in the case where the host galaxy is under
on-going star formation we should consider the possibility that AGN
continuum emission may include a contribution from young stars
(Cid-Fernandes et al. 2004).

Table 1 lists the values of the stellar velocity dispersion and AGN
continuum luminosity (L$_{5100}$) for those 78 BL Lacs which were
successfully modeled with STARLIGHT. The uncertainties in
$\sigma_{\star} $ are the typical errors in the synthesis method,
suggested by Cid Fernandes (2005) based on the S/N at 4020 \AA\ . The
spectral synthesis result for the BL Lac type object J000157.23-103117.3 is
shown in Figure 1, the \emph{top-panel} shows observed spectra (top
black line), host galaxy model (green line), AGN component (red line)
and residuals (bottom black line) are shown. In the \emph{bottom-panel
}, we show details of the modeling around the absorption features.

\begin{figure}
  \includegraphics[width=\columnwidth]{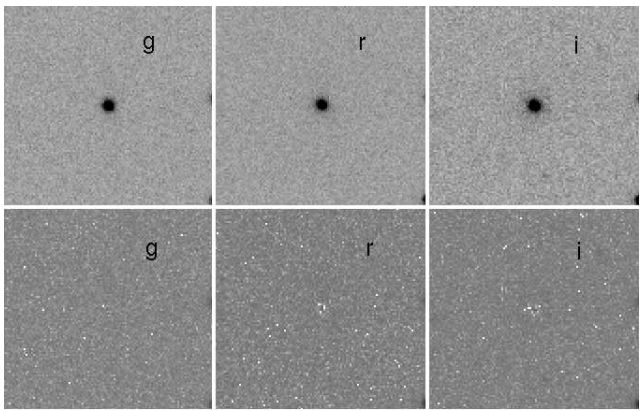}
  \caption{GALFIT decomposition for the BL Lac type
      object J084225.51+025252.7. We show (\emph{top panel}) the 2D
      original images in the $g, r $ and $i$ SDSS-bands, and
      (\emph{bottom panel}) residuals after subtracting the galaxy
      model from the original images. }
\end{figure}

\subsection{Bulge luminosity}

We have modelled the surface brightness profiles of the BL Lac host
galaxies with the two dimensional image decomposition program GALFIT
(Peng et al. 2002). The host galaxy imaging has been retrieved from
the SDSS photometry images in $g, r$ and $i$ bands. Previous studies
(Falomo et al. 2000, Scarpa et al. 2000, Urry et al. 2000, Nilsson et
al. 2003, O'Dowd \& Urry 2005 ) have shown that the luminosity profile
of BL~Lac host galaxies is better represented by an $r^{1/4}$ law (de
Vaucouleurs profile). Therefore, we assume a de Vaucouleurs profile to
model the host galaxy surface brightness in the sample of 78 BL~Lac
objects selected as mentioned in the previous section (3.1). The
initial guesses of the parameters were obtained using Sextractor
(Bertin 1996). Also, we used Sextractor to create mask images for the
fitting. We selected bright non-saturated stars in the BL~Lacs fields
as PSFs. Sky background was fitted first and left fixed during model
fitting. In this way, the total number of parameters is reduced when
de~Vaucouleurs profiles are being computed.

Every BL Lac galaxy is modeled in the \emph{g, r} and
  \emph{i} \ SDSS bands, independently. Then, by comparing the quality
  of the fittings in the 3 bands we decide on the total quality of the
  fitting.  Only the galaxies with good fittings in all three bands
  are considered as part of our analysis.  In Table 1, we list the 71
  BL Lac host galaxies best fitted with a simple de Vaucouleurs
  profile. The bulge magnitude in each band is estimated by using the
  total magnitude of the galaxy and finally the bulge magnitude in
  R-band (M$_{R}$) has been computed using the photometric
  transformation equations found by Jester et al. (2005).  The final
  bulge R-band magnitudes have been corrected for Galactic extinction
  (Schlegel et al. 1998) and we apply the \emph{K-correction} from
  Poggianti (1997).  We include in Table 1 the absolute bulge
  magnitudes for our best fits.  Figure 2 shows the results of the
  structural decomposition for the BL Lac object J084225.51+025252.7.

\begin{table*}\tiny{
\centering
\begin{tabular}{|l|l|r|r|r|r|r|r|r|r|r|}
\hline
  \multicolumn{1}{|c|}{SDSS} &
  \multicolumn{1}{|c|}{Other name} &
  \multicolumn{1}{c|}{$\sigma_{\star}$} &
  \multicolumn{1}{c|}{M$_{BH}$($\sigma_{\star}$)} &
  \multicolumn{1}{c|}{M$_{R}$} &
  \multicolumn{1}{c|}{M$_{BH}$(M$_{R}$)} &
  \multicolumn{1}{c|}{L$_{5GHz}$} &
  \multicolumn{1}{c|}{L$_{5100}$} &
  \multicolumn{1}{c|}{L$_{1keV}$} &
  \multicolumn{1}{c|}{$\nu_{peak}$} &
  \multicolumn{1}{c|}{Class}  \\

  \multicolumn{1}{|c|}{(1)} &
  \multicolumn{1}{|c|}{(2)} &
  \multicolumn{1}{c|}{(3)} &
  \multicolumn{1}{c|}{(4)} &                                                       
  \multicolumn{1}{c|}{(5)} &
  \multicolumn{1}{c|}{(6)} &
  \multicolumn{1}{c|}{(7)} &
  \multicolumn{1}{c|}{(8)} &
  \multicolumn{1}{c|}{(9)} &
  \multicolumn{1}{c|}{(10)} &                                 
  \multicolumn{1}{c|}{(11)}  \\                                          

\hline                                                                                  
                                                                                                   
J000157.23-103117.3 &  NVSS J000157-103118   &  286.26$\pm$ 9.47 &    8.76$\pm$0.10 &    ... &    ... &   31.82 &   28.32 &    25.91 & 13.60       &  LBL  \\
J005620.07-093629.7 &  PMN J0056-0936        &  309.43$\pm$ 4.06 &    8.89$\pm$0.09 &  -22.24$\pm$... &    8.39$\pm$0.17 &   31.51 &   28.55 & $<$26.41 & 16.88       &  HBL  \\
J020106.17+003400.2 &  NVSS J020106+003402   &  208.79$\pm$15.92 &    8.21$\pm$0.15 &  -22.48$\pm$0.01 &    8.51$\pm$0.17 &   31.60 &   28.96 & $<$27.29 & 19.85       &  HBL  \\
J073701.86+284646.0 &                        &  171.92$\pm$11.56 &    7.87$\pm$0.13 &  -21.96$\pm$0.02 &    8.25$\pm$0.17 &   31.57 &   28.89 & $<$26.66 & 17.47       &  HBL  \\
J075437.07+391047.7 &                        &  188.47$\pm$10.15 &    8.03$\pm$0.11 &  -22.04$\pm$0.01 &    8.29$\pm$0.17 &   31.14 &   28.46 & $<$25.40 & 14.20       &  LBL  \\
J075846.99+270515.5 &                        &  198.80$\pm$11.71 &    8.12$\pm$0.12 &  -20.46$\pm$0.01 &    7.50$\pm$0.15 &   31.29 &   27.99 &    25.01 & 12.12       &  LBL  \\
J080018.79+164557.1 &                        &  279.41$\pm$15.16 &    8.71$\pm$0.12 &  -22.80$\pm$0.06 &    8.67$\pm$0.18 &   32.25 &   29.10 &    27.39 & 17.77       &  HBL  \\
J080938.88+345537.1 &  B2 0806+35            &  261.81$\pm$17.39 &    8.60$\pm$0.14 &  -21.47$\pm$... &    8.00$\pm$0.16 &   31.52 &   28.13 & $<$26.22 & \textit{18.29} &  HBL  \\
J084225.51+025252.7 &  NVSS J084225+025251   &  209.30$\pm$11.83 &    8.21$\pm$0.12 &  -22.94$\pm$0.01 &    8.74$\pm$0.17 &   32.00 &   29.28 & $<$26.75 & 16.28       &  HBL  \\
J085036.20+345522.6 &  GB6 J0850+3455        &  244.58$\pm$12.05 &    8.48$\pm$0.11 &  -22.44$\pm$0.01 &    8.49$\pm$0.17 &   31.31 &   28.63 & $<$25.87 & 15.39       &  LBL  \\
J085638.50+014000.6 &  NVSS J085638+014000   &  190.96$\pm$22.76 &    8.05$\pm$0.22 &  -22.65$\pm$0.26 &    8.60$\pm$0.22 &   31.65 &   29.06 &    26.53 & 16.58       &  HBL  \\
J085749.80+013530.3 &  PMN J0857+0135        &  260.24$\pm$ 8.70 &    8.59$\pm$0.09 &  -23.24$\pm$0.01 &    8.89$\pm$0.18 &   32.32 &   29.29 & $<$26.61 & 14.50       &  LBL  \\
J090314.70+405559.8 &                        &  199.99$\pm$21.12 &    8.13$\pm$0.19 &    ... &    ... &   31.52 &   28.31 & $<$26.60 & 17.47       &  HBL  \\
J090953.28+310603.1 &                        &  264.78$\pm$ 8.99 &    8.62$\pm$0.09 &  -23.19$\pm$0.02 &    8.87$\pm$0.18 &   32.24 &   29.09 & $<$27.18 & \textit{17.40} &  HBL  \\
J091045.30+254812.8 &                        &  223.21$\pm$24.20 &    8.32$\pm$0.20 &  -22.47$\pm$0.02 &    8.50$\pm$0.17 &   32.27 &   28.80 &    26.41 & 13.90       &  LBL  \\
J091651.94+523828.3 &                        &  238.27$\pm$11.69 &    8.44$\pm$0.11 &    ... &    ... &   31.92 &   28.77 & $<$26.82 &\textit{ 17.22} &  HBL  \\
J093037.57+495025.6 &  NVSS J093037+495026   &  276.72$\pm$13.08 &    8.70$\pm$0.11 &  -21.61$\pm$0.01 &    8.07$\pm$0.16 &   31.24 &   28.81 & $<$27.48 & \textit{21.13} &  HBL  \\
J094022.44+614826.1 &  NVSS J094022+614825   &  228.58$\pm$17.25 &    8.36$\pm$0.15 &  -21.95$\pm$0.01 &    8.25$\pm$0.16 &   31.10 &   28.55 & $<$26.55 & 18.96       &  HBL  \\
J094542.23+575747.7 &  GB6 J0945+5757        &  256.76$\pm$11.30 &    8.57$\pm$0.10 &  -23.00$\pm$0.01 &    8.77$\pm$0.18 &   32.12 &   29.20 & $<$25.65 & 11.52       &  LBL  \\
J100444.76+375211.9 &  87GB 100148.8+380726  &  236.25$\pm$15.77 &    8.42$\pm$0.13 &  -22.98$\pm$0.01 &    8.76$\pm$0.18 &   32.44 &   29.22 & $<$26.90 & 15.09       &  LBL  \\
J100811.42+470521.4 &  NVSS J100811+470526   &  212.18$\pm$11.26 &    8.23$\pm$0.11 &  -22.08$\pm$0.02 &    8.31$\pm$0.17 &   31.21 &   28.98 & $<$27.70 &\textit{ 19.67} &  HBL  \\
J101706.67+520247.2 &                        &  240.21$\pm$17.79 &    8.45$\pm$0.14 &  -22.37$\pm$0.01 &    8.45$\pm$0.17 &   32.26 &   28.89 & $<$26.00 & 12.41       &  LBL  \\
J102013.76+625010.1 &                        &  230.21$\pm$20.52 &    8.38$\pm$0.17 &    ... &    ... &   31.82 &   28.60 & $<$26.14 & 14.50       &  LBL  \\
J102523.04+040229.0 &  PMN J1025+0402        &  171.02$\pm$14.44 &    7.86$\pm$0.16 &  -21.69$\pm$0.01 &    8.12$\pm$0.16 &   31.82 &   28.49 & $<$26.30 & 15.09       &  LBL  \\
J103317.94+422236.4 &  GB6 J1033+4222        &  245.46$\pm$24.05 &    8.49$\pm$0.18 &  -22.03$\pm$0.01 &    8.28$\pm$0.17 &   31.67 &   28.34 & $<$25.74 & 13.60       &  LBL  \\
J103346.39+370825.1 &                        &  311.21$\pm$15.68 &    8.90$\pm$0.12 &    ... &    ... &   31.45 &   29.06 & $<$26.83 & 18.66       &  HBL  \\
J104029.01+094754.2 &                        &  315.35$\pm$11.63 &    8.93$\pm$0.11 &  -22.44$\pm$0.01 &    8.49$\pm$0.17 &   31.74 &   28.94 &    27.17 & 18.96       &  HBL  \\
J105606.61+025213.4 &  NVSS J105606+025227   &  184.21$\pm$19.86 &    7.99$\pm$0.20 &  -21.69$\pm$0.25 &    8.12$\pm$0.21 &   30.93 &   28.59 & $<$27.43 & 22.83       &  HBL  \\
J110222.94+380122.5 &                        &  262.50$\pm$21.49 &    8.60$\pm$0.16 &  -22.29$\pm$0.02 &    8.42$\pm$0.17 &   32.04 &   28.86 &    26.22 & 13.90       &  LBL  \\
J110356.14+002236.3 &                        &  249.22$\pm$23.42 &    8.51$\pm$0.18 &  -22.42$\pm$0.02 &    8.48$\pm$0.17 &   32.08 &   28.79 &    25.96 & 12.71       &  LBL  \\
J112059.74+014456.9 &                        &  210.44$\pm$22.18 &    8.22$\pm$0.19 &  -22.19$\pm$0.12 &    8.36$\pm$0.18 &   32.02 &   28.71 & $<$26.83 & 16.28       &  HBL  \\
J114023.48+152809.7 &                        &  262.01$\pm$13.73 &    8.60$\pm$0.12 &  -22.88$\pm$... &    8.71$\pm$0.17 &   32.07 &   29.21 & $<$27.24 & 17.77       &  HBL  \\
J114535.10-034001.4 &                        &  186.99$\pm$19.43 &    8.01$\pm$0.19 &  -21.66$\pm$0.03 &    8.10$\pm$0.16 &   30.98 &   28.04 & $<$26.81 & 20.45       &  HBL  \\
J115404.55-001009.8 &                        &  209.72$\pm$12.32 &    8.21$\pm$0.12 &  -22.17$\pm$0.01 &    8.35$\pm$0.17 &   31.30 &   28.65 & $<$26.95 & 19.55       &  HBL  \\
J120208.65+444422.4 &  B3 1159+450           &  262.82$\pm$ 7.52 &    8.61$\pm$0.09 &  -22.75$\pm$0.09 &    8.65$\pm$0.18 &   32.18 &   29.32 &    25.84 & \textit{15.92} &  LBL  \\
J120303.50+603119.1 &  GB6 J1203+6031        &  195.69$\pm$12.86 &    8.09$\pm$0.13 &  -21.46$\pm$... &    8.00$\pm$0.16 &   31.35 &   28.92 & $<$25.14 & 12.41       &  LBL  \\
J120412.11+114555.4 &                        &  251.26$\pm$19.47 &    8.53$\pm$0.15 &  -22.90$\pm$0.10 &    8.72$\pm$0.18 &   31.70 &   29.44 & $<$26.76 & 17.47       &  HBL  \\
J122300.30+515313.9 &                        &  187.26$\pm$21.14 &    8.02$\pm$0.21 &  -21.81$\pm$0.02 &    8.18$\pm$0.16 &   31.16 &   28.85 &    26.13 & 16.88       &  HBL  \\
J122809.13-022136.1 &                        &  177.01$\pm$22.48 &    7.92$\pm$0.23 &  -21.74$\pm$0.02 &    8.14$\pm$0.16 &   30.69 &   28.39 & $<$26.82 & 21.64       &  HBL  \\
J123123.90+142124.4 &  GB6 J1231+1421        &  312.09$\pm$10.34 &    8.91$\pm$0.10 &  -23.00$\pm$... &    8.77$\pm$0.18 &   32.00 &   29.39 &    26.48 & \textit{14.91} &  LBL  \\
J123623.01+390001.0 &  GB6 J1236+3859        &  253.51$\pm$14.53 &    8.54$\pm$0.12 &  -23.22$\pm$0.01 &    8.88$\pm$0.18 &   32.25 &   29.23 & $<$26.51 & \textit{16.61} &  LBL  \\
J123739.08+625842.8 &                        &  215.18$\pm$11.40 &    8.26$\pm$0.11 &  -22.68$\pm$0.02 &    8.61$\pm$0.17 &   31.54 &   29.00 & $<$27.06 & \textit{15.98} &  HBL  \\
J123831.24+540651.8 &                        &  264.56$\pm$20.68 &    8.62$\pm$0.15 &  -22.17$\pm$0.01 &    8.36$\pm$0.17 &   31.70 &   28.71 &    25.46 & 12.41       &  LBL  \\
J124834.30+512807.8 &  87GB 124615.8+514411  &  263.17$\pm$ 8.24 &    8.61$\pm$0.09 &  -23.39$\pm$0.01 &    8.96$\pm$0.18 &   32.56 &   29.44 & $<$26.38 & 12.41       &  LBL  \\
J125347.00+032630.3 &                        &  218.17$\pm$17.82 &    8.28$\pm$0.16 &  -21.81$\pm$0.01 &    8.18$\pm$0.16 &   31.02 &   28.18 & $<$25.53 & 15.09       &  LBL  \\
J131330.12+020105.9 &                        &  254.58$\pm$13.41 &    8.55$\pm$0.11 &  -22.72$\pm$0.02 &    8.63$\pm$0.17 &   32.33 &   29.06 & $<$26.54 & 14.20       &  LBL  \\
J132301.00+043951.3 &                        &  280.33$\pm$20.30 &    8.72$\pm$0.15 &  -22.28$\pm$0.01 &    8.41$\pm$0.17 &   31.67 &   28.50 & $<$26.90 & 18.07       &  HBL  \\
J132617.70+122958.7 &                        &  269.02$\pm$16.43 &    8.65$\pm$0.13 &  -21.95$\pm$0.01 &    8.24$\pm$0.16 &   31.59 &   28.49 & $<$26.89 & \textit{16.32} &  HBL  \\
J133102.91+565541.8 &                        &  224.27$\pm$22.19 &    8.33$\pm$0.18 &  -21.67$\pm$0.03 &    8.10$\pm$0.16 &   30.88 &   28.46 & $<$26.27 & 18.66       &  HBL  \\
J133105.70-002221.2 &                        &  181.24$\pm$19.87 &    7.96$\pm$0.20 &  -21.61$\pm$0.01 &    8.07$\pm$0.16 &   31.21 &   28.41 & $<$26.00 & 16.28       &  HBL  \\
J134105.10+395945.4 &                        &  258.51$\pm$14.22 &    8.58$\pm$0.12 &  -22.18$\pm$0.01 &    8.36$\pm$0.17 &   31.59 &   28.19 & $<$26.82 & \textit{20.06} &  HBL  \\
J134136.23+551437.9 &  87GB 133948.0+552941  &  222.53$\pm$13.45 &    8.32$\pm$0.12 &  -21.90$\pm$0.02 &    8.22$\pm$0.16 &   31.65 &   28.65 & $<$26.16 & 15.39       &  LBL  \\
J134502.30+553914.2 &                        &  229.72$\pm$20.08 &    8.37$\pm$0.16 &  -21.52$\pm$0.01 &    8.03$\pm$0.16 &   29.94 &   28.39 &    24.96 & 17.17       &  HBL  \\
J134633.98+244058.4 &  NVSS J134634+244100   &  209.73$\pm$15.24 &    8.21$\pm$0.14 &  -22.24$\pm$0.01 &    8.39$\pm$0.17 &   31.37 &   28.34 &    25.16 & 12.41       &  LBL  \\
J135314.08+374113.9 &  87GB 135107.4+375518  &  289.96$\pm$18.72 &    8.78$\pm$0.14 &  -22.75$\pm$0.26 &    8.64$\pm$0.22 &   31.70 &   28.76 & $<$26.13 & 14.79       &  LBL  \\
J140121.13+520928.9 &                        &  227.32$\pm$20.68 &    8.35$\pm$0.17 &  -22.69$\pm$0.02 &    8.61$\pm$0.17 &   32.07 &   29.06 & $<$26.87 & 16.28       &  HBL  \\
J140330.85+360651.1 &                        &  122.47$\pm$21.26 &    7.27$\pm$0.32 &  -21.20$\pm$0.02 &    7.87$\pm$0.16 &   30.78 &   28.50 &    25.47 & 15.98       &  HBL  \\
J140923.50+593940.7 &                        &  258.21$\pm$17.87 &    8.58$\pm$0.14 &  -23.45$\pm$0.01 &    9.00$\pm$0.18 &   32.27 &   29.11 & $<$27.26 & \textit{16.63} &  HBL  \\
J141030.84+610012.8 &                        &  300.91$\pm$23.53 &    8.84$\pm$0.16 &  -22.47$\pm$0.02 &    8.50$\pm$0.17 &   31.40 &   28.62 & $<$27.25 & \textit{20.25} &  HBL  \\
J142832.60+424021.0 &  87GB 142634.5+425353  &  260.35$\pm$14.53 &    8.59$\pm$0.12 &  -22.14$\pm$... &    8.34$\pm$0.17 &   31.35 &   28.85 & $<$27.40 & \textit{18.55} &  HBL  \\
J144248.28+120040.2 &                        &  310.80$\pm$13.44 &    8.90$\pm$0.11 &  -22.00$\pm$0.01 &    8.27$\pm$0.17 &   31.77 &   29.02 & $<$26.98 & \textit{16.45} &  HBL  \\
J145326.52+545322.4 &                        &  211.80$\pm$22.71 &    8.23$\pm$0.20 &  -21.45$\pm$0.01 &    8.00$\pm$0.16 &   29.62 &   28.14 & $<$24.78 & 17.77       &  HBL  \\
J153447.21+371554.5 &  87GB 153254.4+372523  &  154.11$\pm$12.19 &    7.67$\pm$0.15 &  -21.48$\pm$0.02 &    8.01$\pm$0.16 &   31.15 &   28.70 & $<$25.35 & \textit{14.26} &  LBL  \\
J160519.04+542059.9 &                        &  165.26$\pm$18.65 &    7.80$\pm$0.21 &  -21.02$\pm$0.03 &    7.78$\pm$0.16 &   30.90 &   28.30 & $<$26.93 & 21.04       &  HBL  \\
J161541.21+471111.7 &                        &  218.22$\pm$13.65 &    8.28$\pm$0.13 &  -22.13$\pm$... &    8.34$\pm$0.17 &   32.09 &   28.81 &    25.43 & 10.63       &  LBL  \\
J161706.32+410647.0 &                        &  184.60$\pm$ 9.41 &    7.99$\pm$0.11 &  -22.55$\pm$0.01 &    8.54$\pm$0.17 &   32.16 &   29.10 & $<$26.71 & \textit{14.41} &  LBL  \\
J162839.03+252755.9 &                        &  311.76$\pm$19.34 &    8.91$\pm$0.14 &  -22.38$\pm$0.02 &    8.46$\pm$0.17 &   31.90 &   28.72 & $<$26.65 & 16.28       &  HBL  \\
J163709.50+432600.3 &  NVSS J163709+432600   &  299.05$\pm$11.90 &    8.83$\pm$0.11 &  -23.09$\pm$0.01 &    8.81$\pm$0.18 &   32.29 &   29.18 & $<$25.94 & 11.82       &  LBL  \\
J163726.66+454749.0 &  B3 1635+458           &  245.11$\pm$23.43 &    8.49$\pm$0.18 &  -22.10$\pm$0.01 &    8.32$\pm$0.17 &   31.28 &   28.44 & $<$26.11 & 16.58       &  HBL  \\
J164419.97+454644.4 &  B3 1642+458           &  304.30$\pm$17.31 &    8.86$\pm$0.13 &  -22.24$\pm$0.23 &    8.39$\pm$0.20 &   32.27 &   28.72 & $<$26.63 & 14.79       &  LBL  \\
J165109.19+421253.4 &                        &  175.21$\pm$18.60 &    7.90$\pm$0.20 &  -21.35$\pm$0.01 &    7.94$\pm$0.16 &   31.60 &   28.54 &    25.60 &\textit{ 17.82} &  LBL  \\
J171427.40+560156.0 &                        &  268.51$\pm$23.41 &    8.64$\pm$0.17 &    ... &    ... &   31.53 &   28.70 & $<$26.59 & 17.47       &  HBL  \\
J172918.78+525559.2 &                        &  175.31$\pm$14.80 &    7.90$\pm$0.16 &  -22.85$\pm$0.36 &    8.69$\pm$0.25 &   31.91 &   29.08 & $<$26.79 & 16.58       &  HBL  \\
J211037.75-072412.9 &  NVSS J211038-072408   &  254.84$\pm$17.50 &    8.55$\pm$0.14 &  -21.79$\pm$0.03 &    8.17$\pm$0.16 &   30.59 &   28.47 &    25.59 & 17.17       &  HBL  \\
J211611.89-062830.4 &                        &  293.34$\pm$13.26 &    8.80$\pm$0.11 &  -22.19$\pm$0.01 &    8.37$\pm$0.17 &   31.61 &   28.93 &    26.13 & 15.39       &  LBL  \\
J221109.88-002327.4 &                        &  273.75$\pm$24.05 &    8.68$\pm$0.17 &    ... &    ... &   32.40 &   28.86 &    26.70 & 14.50       &  LBL  \\
J222944.18-003426.6 &                        &  260.27$\pm$12.75 &    8.59$\pm$0.11 &  -22.90$\pm$0.05 &    8.72$\pm$0.18 &   30.93 &   28.65 &    26.82 & 20.74       &  HBL  \\
J235604.01-002353.7 &                        &  182.99$\pm$20.81 &    7.97$\pm$0.21 &  -21.83$\pm$0.01 &    8.19$\pm$0.16 &   31.12 &   28.68 & $<$26.20 & 17.47       &  HBL  \\

\hline\end{tabular}
}
\caption{Column description: (1) SDSS name, (2) alias  given by radio surveys, (3) stellar velocity dispersions in units (km s$^{-1}$), (4) log of black hole mass estimated from stellar velocity dispersion in units of (M$_\odot$), (5) R-band bulge absolute magnitude, (6) log of black hole mass estimated from R-band bulge luminosity in units of (M$_\odot$), (7) log of radio luminosity at  5GHz, (8) log of nuclear optical continuum luminosity at 5100 \AA~,  (9) log of X-ray luminosity at 1 keV, (10) log of synchrotron peak frequency  in (Hz) and (11) the classification of BL Lacs  based on the X-ray to radio flux ratio, following Padovani \& Giommi (1995). Note: All luminosities  in this table are  in units of erg s$^{-1}$ Hz$^{-1}$ and synchrotron peak frequencies in italic font have been taken from Nieppola et al. (2006)}
\end{table*}

\section{BLACK HOLE MASSES}

\begin{figure}
\includegraphics[width=\columnwidth]{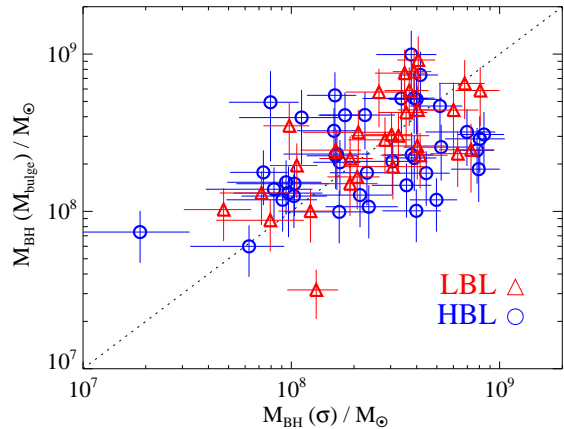}
\caption{Comparison of black hole mass estimates of
    71 BL Lac type objects from the stellar velocity dispersion and
    from the R-band bulge magnitude.  There is no significant
    difference in the black hole mass range between LBL
    (\emph{open-triangles}) and HBL (\emph{open-circles}) objects. The
    \emph{dashed line} respresents the one-to-one correspondence.} 
\end{figure}

Following Tremaine et al. (2002) and McLure and Dunlop
  (2002), the expressions to estimate black hole masses can be
  expressed in the following forms:

\begin{equation}\label{eq:msigma}
  \log M_{BH}(\sigma_{\star})  =  4.02 (\pm 0.32)\log \left( \frac{\sigma_\star} {\sigma_{0}} \right) + 8.13 (\pm 0.06)
\end{equation}

\begin{equation}\label{eq:mbulge}
  \log M_{BH}(M_{R})  = -0.50 (\pm 0.02)~M_{R}  + 2.73 (\pm 0.48)
\end{equation}
where $\sigma_{\star}$ is the stellar velocity dispersion,
  $\sigma_{0}=$ 200 km s$^{-1}$ and M$_{R}$ is the absolute bulge
  magnitude in R band (corrected to our adopted cosmology). The
black hole masses derived from $\sigma_{\star}$ and M$_{R}$ are
designated as M$_{BH} (\sigma_{\star})$ and M$_{BH} (M_{R})$,
respectively in columns 4 and 6 of Table 1.  As can be seen from
Figure 3, there is a generally good agreement between black hole
masses estimated by both methods.  The average values derived from
$\sigma_{\star}$ and $M_{R}$ are $\langle log~M_{BH}\rangle
_{\sigma_{\star}} = 8.38 \pm 0.04 $ and $\langle
  log~M_{BH}\rangle _{M_{R}} = 8.38 \pm 0.03$, respectively, with an
  average difference of $\langle \Delta log~M_{BH}\rangle = -0.008 \pm
  0.038$ . This result gives us additional confidence in the
reliability of our black hole mass estimations. Since the relation
M$_{BH}- \sigma_\star$ suffers of less intrinsic dispersion, is the
most accurate (Vestergaard 2009) and $\sigma_{\star}$ measurements are
not affected by potential beaming effects, in the following we focus
our analysis on black hole masses estimated via stellar velocity
dispersion.

Figure 4 shows the distribution of black hole masses estimated for HBL
and LBL objects.  As can be seen, the ranges of black hole mass in
both populations are consistent with each other. The
Kolmogorov-Smirnov test shows that there is no a significant
difference between the distributions of black hole mass for HBL and
LBL objects, confirming results in previous BL Lacs studies(Falomo et
al. 2002; 2003, Woo et al. 2005)

\begin{figure}
\includegraphics[width=\columnwidth]{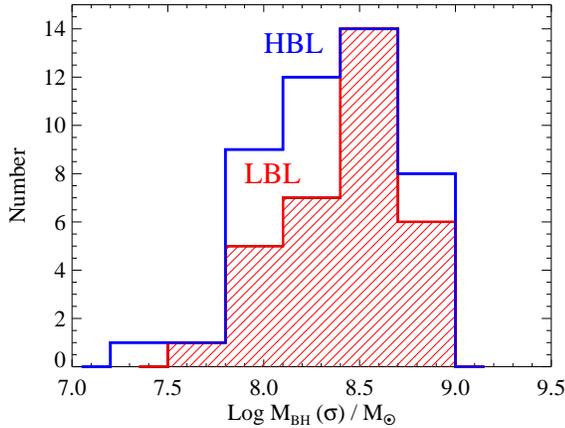}
\caption{M$_{BH}$ distribution for 78 BL Lac type objects.  There is
  no a significant difference in the black hole mass distribution
  between HBL (\emph{blue}) and LBL (\emph{red}) objects.}
\end{figure}

\section{THE BLACK HOLE MASS-  LUMINOSITY RELATIONSHIP}

\begin{figure}
\includegraphics[width=\columnwidth]{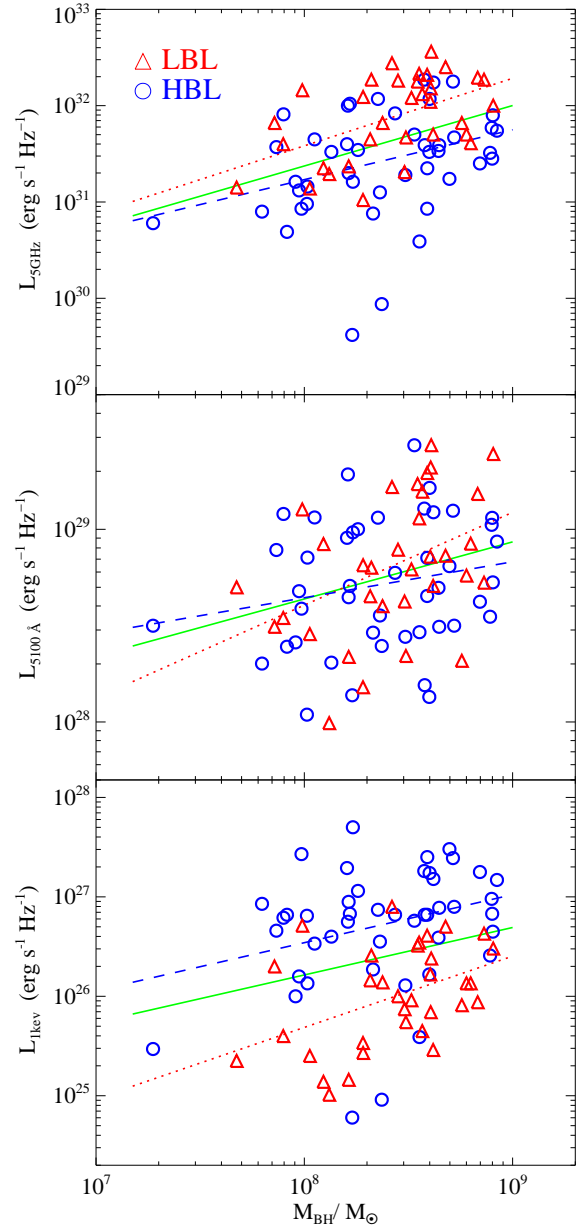}
\caption{ Correlation between black hole mass and luminosity at $5GHz$
  (\emph{top panel}), $5100\AA~$ (\emph{middle panel}) and $1keV$
  (\emph{bottom panel}). HBL and LBL are shown by \emph{open circles}
  and \emph{open triangles}, respectively. The \emph{solid line} gives
  the ordinary least-squares fit to the data, whereas fitted relations
  for LBL and HBL separately are shown as \emph{dotted} and
  \emph{dashed} lines, respectively. See Table 2 for the correlation
  and best-fit parameters. }\label{fig:mass_luminosity}
\end{figure}

In Figure 5 we plot our black hole mass estimations versus
multiwavelength luminosities. Luminosities at 5 GHz and 1keV have been
taken from Plotkin et al. (2008). Disentangling the AGN component from
the host galaxy contribution as described in section 3.1 allowed us to
recover the optical nuclear continuum luminosity at 5100~\AA. The
solid green lines in the panels of Figure 5 represent our fit to the
data. Treating M$_{BH}$ as the independent variable, an ordinary
least-squares fit yields

\begin{equation}
L_{5GHz} \propto M_{BH} ^{ \ 0.63\pm0.16}  
\end{equation}
\begin{equation}
L_{5100 \AA} \propto M_{BH} ^{ \ 0.30\pm0.11}
\end{equation}
\begin{equation}
L_{1keV} \propto M_{BH} ^{ \ 0.48\pm0.21}
\end{equation}

The Spearman rank correlation coefficients listed in Table 2, confirm
a positive correlation for the above relations at high significance
level.  These relations are fits for the combined HBL and LBL
populations.  However, the best-fit parameters and correlation
strengths are slightly changed when HBLs and LBLs are considered
separately.  The fitted relations for LBL and HBL are shown in Figure
5 as dotted and dashed lines, respectively.  Table 3 lists the
best-fit parameters for each BL Lac population. It is worth noticing
that the slopes derived for the LBL population are somewhat steeper
than those derived for the HBL population.

The correlation between black hole mass and X-ray luminosity becomes
stronger and more significant when computed separately for LBL and HBL
populations (see Table 2). This arises naturally if we bear in mind
that HBL and LBL are by definition two different populations in the
X-ray regime.  The Spearman rank correlation test provides evidence of
a positive correlation between M$_{BH}$ and L$_{5GHz}$, L$_{5100}$ and
L$_{1keV}$ at a confidence level $> 98\%$ when restricted only to LBL.
We note that when considering exclusively HBLs, the correlations
between M$_{BH}$ and L$_{5GHz}$, L$_{1keV}$ are weaker, though still
significant ($> 95\%$). However, the Spearman rank correlation test
between black hole mass and optical luminosity for HBL provides only
evidence of a trend (at 80$\%$ confidence), this may betray the
presence of a stellar contribution in the optical continuum luminosity
of HBLs. Whether the stellar component differs significantly between
LBL and HBL will be addressed in a forthcoming paper. As mentioned
before, the correlations described above have been computed using the
black hole mass derived from stellar velocity dispersion. However, the
best-fit parameters are virtually unchanged when the estimations of
black hole mass via bulge magnitude are considered.

\begin{table}
  \caption[]{Spearmank's  correlation coefficients between black hole mass and multiwavelength luminosities  for  BLLacs  in our sample. Partial correlation methods have been performed for correlations of the  luminosity-luminosity type, in order to remove the common dependence with redshift. We consider a correlation to be significant if the probability that the correlation is given by chance is  $P \leq 5 \times 10^{-2}$  or in other words  at a significance level of 95\%}
\begin{tabular}{cccccccccccccc}
 \hline \hline

   &  & \multicolumn{2}{c}{\textcolor{green}{All}} &  &  \multicolumn{2}{c}{\textcolor{red}{LBL}} & & \multicolumn{2}{c}{\textcolor{blue}{HBL}}\\
 \cline{3-4}\cline{6-7}\cline{9-10} \smallskip
  A1 & A2   &   $\rho$ & $P$ &  &  $\rho$ & $P$ & & $\rho$ & $P$ \\
 \hline\hline
$M_{\rm BH}$ & $ L_{\rm 5GHz}$            &  0.4 & 2$\times 10^{-4}$ & &0.5 &5$\times 10^{-3}$& &0.4 & 1$\times 10^{-2}$\\
$M_{\rm BH}$ & $ L_{\rm 5100\AA}$       &  0.3 & 9$\times  10^{-3}$& &0.5 &8$\times 10^{-3}$& &0.2 & 2$\times 10^{-1}$\\
$M_{\rm BH}$ & $ L_{\rm 1keV}$             &  0.2 & 7$\times 10^{-2}$ & &0.4 &2$\times 10^{-2}$& &0.3 & 4$\times 10^{-2}$\\
$ L_{\rm 5GHz}$ & $ L_{\rm 5100\AA}$  &  0.6 & 3$\times 10^{-9}$ & & 0.5 & 6$\times 10^{-3}$& &0.7& 2$\times 10^{-8}$\\
$ L_{\rm 5GHz}$ & $ L_{\rm 1keV}$        &  0.1 & 7$\times 10^{-1}$  & &0.4 & 1$\times 10^{-2}$& &0.4 & 8$\times 10^{-3}$\\
 \hline

\end{tabular}

\end{table}

\begin{table}
  \caption[]{Best fit parameters  for the M$_{BH}$-luminosity and luminosity-luminosity  relations shown in Figure 5 and 6, respectively.}
  
\begin{tabular}{cccccccccccccc}
 \hline \hline

      & \textcolor{green}{All}   &  \textcolor{red}{LBL}  & \textcolor{blue}{HBL}\\
      & $x$   &  $x$  & $x$\\
 \hline\hline
 $L_{\rm 5GHz} \propto M_{\rm BH}^x$     &$0.63\pm0.16$ & $0.70\pm0.21$  &  $0.51\pm0.19$  \\
 $L_{\rm 5100\AA} \propto M_{\rm BH}^x$  &$0.30\pm0.11$ & $0.48\pm0.19$  &  $0.18\pm0.14$  \\
 $L_{\rm 1keV} \propto M_{\rm BH}^x$     &$0.48\pm0.22$ & $0.71\pm0.26$  &  $0.48\pm0.21$  \\
 $L_{\rm 5GHz} \propto L_{\rm 5100}^x$ &$1.00\pm0.13$ & $0.86\pm0.14$  &  $1.02\pm0.18$  \\
 $L_{\rm 5GHz} \propto L_{\rm 1keV}^x$    &$0.24\pm0.09$ & $0.63\pm0.10$  &  $0.60\pm0.10$  \\
 \hline

\end{tabular}

\end{table}

In search of some physical insight into the black hole mass -
luminosity relation, we explore correlations between X-ray/optical and
radio emission. For luminosity-luminosity correlations we have taken
into account the common dependence with the redshift by using partial
correlation analysis. The results of our partial correlation analysis
between luminosities are given in Table 2.  Radio and optical
continuum luminosities are shown in the top panel of figure 6, where a
significant correlation between L$_{5GHz}$ and L$_{5100}$ is found at
very high confidence level with a correlation coefficient $\rho= 0.6$
and the probability that such correlation is not due to chance is
$>99.9\%)$ . This suggests that the emission in the two bands has a
similar origin, more specifically non-thermal radiation from the
jets. The best fit to the combined sample is shown as a solid line in
the top panel of Figure 6, whereas dotted and dashed lines represent
the best fits for the LBL and HBL populations.

Nevertheless, when attempting to compare the distributions of
L$_{5GHz}$ and X-ray luminosities, it turns out that we do not find
evidence of significant correlation when the dependence of redshift is
removed (see bottom panel in Figure 6).  However, when HBL and LBL are
considered separately, the Spearman rank correlation strength between
luminosities becomes higher and significant ($> 98\%$). This comes
naturally, considering that HBL and LBL form two different populations
in the X-ray regime. Table 1 lists correlation and best-fit parameters
for each population. We stress that when computing correlations with
the X-ray luminosities, we use upper limits for those sources non
detected in Xrays.  It is conceivable that real values would produce
an even higher separation between HBL and LBL populations, which would
not change the significance of correlations computed.

\begin{figure}
   \includegraphics[width=\columnwidth]{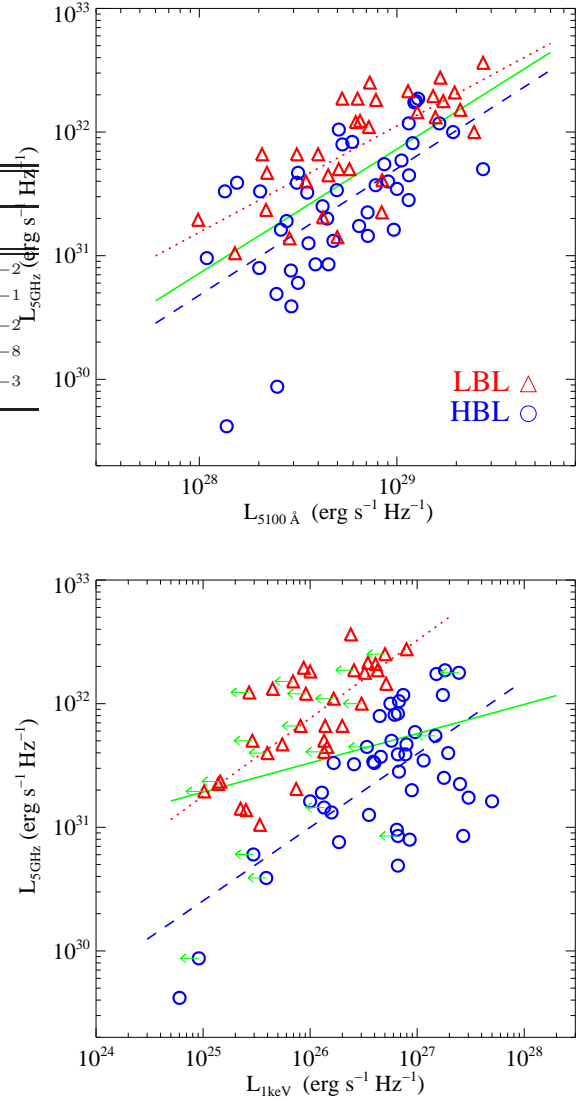}
   \caption{ Correlation between radio luminosity at 5GHz and optical
     continuum (\emph{left panel}) and Xray (\emph{right panel})
     emission.The correlation between X-rays and radio emission only
     holds when considering HBL and LBL populations separately. The
     \emph{solid line} gives the ordinary least squares fit to the
     data, where fitted relations for LBL and HBL populations are
     shown as \emph{dotted} and \emph{dashed} lines,
     respectively. Best-fit parameters and correlation coefficients
     among luminosities are listed in Table 3.  }
\end{figure}

\section{Discussion}

In this work we have measured stellar velocity dispersions and bulge
luminosities for BL Lac objects identified in the SDSS and FIRST
surveys, enabling us to estimate masses for the black holes in our
sample.  The range of black hole masses computed from stellar velocity
dispersion is in substantial agreement with those estimated from
R-band bulge magnitude, the average difference of black hole
  masses estimated by both methods is $\langle \Delta
  log~M_{BH}\rangle = -0.008 \pm 0.038$ . Within the sample of
BL~Lacs considered, LBL and HBL sources are indistinguishable in terms
of black hole mass.

We found that in our sample of BL~Lacs, the black hole mass
correlates at high significance level with radio (5GHz), optical
nuclear (5100 \AA), and X-ray (1keV) luminosities.  Radio luminosity
at 5GHz correlates tightly with the nuclear optical luminosity and
with X-ray luminosity when correlation coefficients are computed for
LBL and HBL, separately, implying that the emission in the optical and
X-ray bands share a similar origin, more specifically non-thermal
emission from the jets.

We note that the strong correlation between black hole mass and radio
luminosity found in this study is in disagreement with the results of
Woo et al (2005,) who found no evidence of such a correlation using a
sample of BL Lacs. The range of black hole masses found in this work
are similar than those found in Woo et al. (2005) and other previous
works (Falomo et al. 2003; Dunlop et al. 2003).  Then, these
apparently conflicting results may be attributable to the literature
compilation done by Woo et al. (2005) where the measurements have not
been obtained in a homogeneous way. Whereas in this work all the black
hole masses have been estimated in a uniform fashion, data taken with
the same instruments. Besides, various studies have found a roughly
linear correlation between black hole mass and radio luminosities
(e.g. Falomo et al. 2003, Dunlop et al. 2003, Arshakian et al. 2005),
although consistent with our study they report a slope higher than
ours.  This apparent disagreement may be related to the mixed
population of radio sources involved in those studies and also due to
the different luminosity coverage.  The black hole masses used in the
previous studies are compiled from literature and mainly computed from
the broad emission lines, which might introduce potential
uncertainties regarding the location, geometry and non-virialization
effects in the BLR (Arshakian et al. 2010, Le\'on-Tavares et al. 2010,
Ili\'c et al. 2009). The advantage of this work over the previous
ones, is that our measurements have been performed in a uniform
fashion and the 78 black hole masses estimated from the stellar
velocity dispersion are free of the potential uncertainties mentioned
above.

 \begin{figure}
\includegraphics[width=\columnwidth]{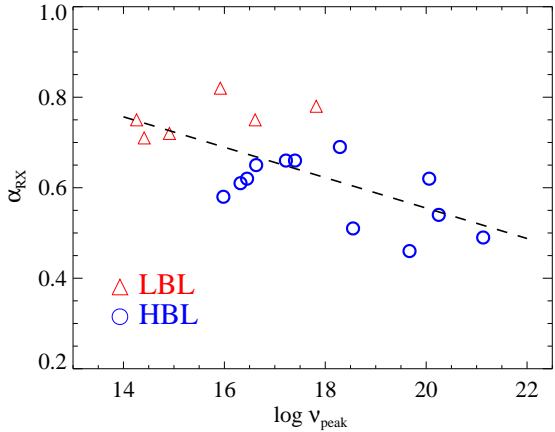}
\caption{The distribution of radio-X-ray spectral index
  ($\alpha_{rx}$) and synchrotron peak frequency ($\nu_{peak}$) for
  BLLacs which SED has been succesfully modelled in Nieppola et
  al. 2006. The \emph{dashed line} gives the ordinary least squares
  fit to the data.  }
 \end{figure}

 Since it is well established that the synchrotron emission in BLLacs
 (and in blazars in general) is affected by relativistic beaming
 effects due to a fast jet aligned close to our line of sight
 (Blandford \& Konigl 1979), then we must expect the emission in
 BL Lacs to be Doppler boosted. However, we should be aware that
 luminosities in BLLacs might not exclusively depend on Doppler
 boosting effects but on intrinsic physics conditions.

 As a consequence, the correlation M$_{BH} - luminosity$ found in this
 work, might be understood either as an evidence of observed
 dependence between black hole mass and Doppler boosting effects
 (Arshakian et al. 2005, Valtaoja et al. 2008, Torrealba et al. 2008)
 or as a relationship between black hole mass and intrinsic physics
 conditions in each source (i.e. magnetic field strength, jet
 composition, etc).  Unfortunately we can not test \emph{in situ} the
 correlation between black hole mass and Doppler factor with our
 sample of BL Lacs, due to the fact that computation of variability
 Doppler factors are observationally expensive, requiring long-term
 monitoring.  However, we may attempt to disentangle the population of
 BLLacs for which luminosity is most likely Doppler boosted, as
 follows.

\begin{figure}
 \includegraphics[width=\columnwidth]{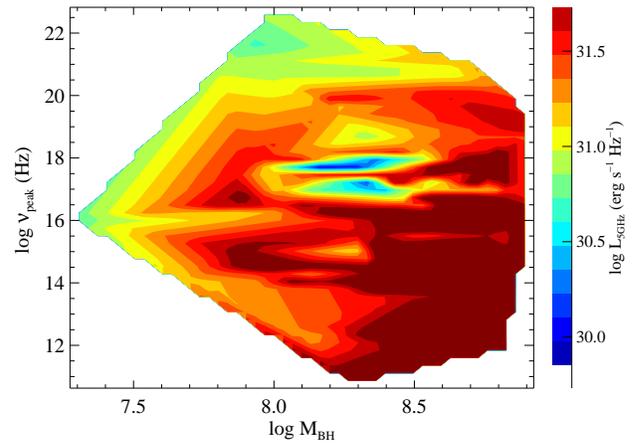}
 \caption{ Black-hole mass - synchrotron peak frequency plane, where
   the color scale corresponds to the intensity of radio emission,
   darker colors corresponding to brightest radio-jets . The
   \emph{region} located at the bottom-right of the plane is
   exclusively populated by the heaviest black-holes ($M_{BH} >
   10^{8.25}$), brightest radio-jets ($L_{5GHz} > 10^{31.5}$) and the
   lowest synchrotron peak frequencies ($\nu_{peak} <
   10^{16}$). Sources contained in this \emph{beamed-region} fullfill
   the criteria for which broad band emission is dominated by Doppler
   boosting effects. }
 \end{figure}

 Starting from the hypothesis that emission strength in BL Lacs
 depends primarily on Doppler boosting effects which are characterized
 by the Doppler factor ($D$), then derived expressions (5-7) can be
 seen as L$(D) \propto M_{BH}$. Furthermore, it is well established
 that emission from the relativistic beamed jet is responsible for the
 lower energy-peak in the BL Lacs SED and its position on the
 log$\nu$-axis is defined as the synchrotron-peak frequency
 ($\nu_{peak}$). Using a sample of 89 blazars, Nieppola et al. (2008)
 found an inverse dependence between the Doppler boosting factors and
 peak frequencies ($D \propto \nu_{peak}^{-1}$). Such a relation holds
 for $\nu_{peak} < 10^{16}$ (Hz), above which Nieppola et al. (2008)
 argued the D$ \sim 1$. Based on this, we may assume that Doppler
 boosted emission follow the relation L$(D) \propto
 \nu_{peak}^{-1}$. According to these expressions, we should expect
 the dependence between black-hole mass and synchrotron peak frequency
 to be inversely proportionally (M$_{BH} \propto \nu_{peak}^{-1}$) for
 BL Lacs where emission is dominated by Doppler boosting effects.

In order to find whether BLLacs in our sample fulfill
  the relations mentioned above, we compile synchrotron peak
  frequencies estimates for our sample. SEDs for 18 BLLacs contained
  in this study have been constructed and successfully modeled in
  Nieppola et al. (2006). The distribution of peak frequencies with
  their respective broad band spectral indexes ($\alpha_{RX}$) is
  shown in Figure 7. Information about the broad band spectral indexes
  has been taken from Table 6 in Plotkin et al. (2008) . A common
  least-square fit to the data yields, 

\begin{equation}
\alpha_{RX} = (1.23 \pm 0.04) +  log (\nu_{peak}) \times (-0.03 \pm 0.01)
\end{equation}

This expression allows us to compute the peak frequencies for the rest
of our sample. The compiled and computed synchrotron peak frequencies
are listed in Table 1. Figure 8 shows the $M_{BH}-\nu_{peak}$ plane
where the color scale corresponds to the radio luminosity, darker
colors corresponding to brightest radio jets. We note that there is
very well defined region located at the bottom-right part of the
plane. This \emph{region} is exclusively populated by the heaviest
black-holes($M_{BH} > 10^{8.25}$), brightest radio-jets ($L_{5GHz} >
10^{31.5}$) and the lowest peak frequencies ($\nu_{peak} < 10^{16}$)
and in the following we refer to it as the \emph{beamed region}. 32
out of 78 ($40\%$) BL Lacs in our sample (mostly classified as LBL)
are enclosed in this \emph{beamed region}, fullfilling the criteria
for which the broad-band emission is dominated by Doppler boosting
effects.

Doppler boosting effects in a relativistic jet are often characterized
by the Doppler factor, which is a function of the jet viewing angle
and jet speed.  Since the jet viewing angle does not play a major role
in determining the SED shape in BL Lac type objects (Padovani \&
Giommi 1995), we may assume that in our entire sample of BLLacs (LBL
and HBL), the Doppler boosted luminosities will not strongly depend on
the orientation angle but mainly on the speed of the jet. Hence, the
correlation $M_{BH}-luminosity$ for BL Lacs inside the \emph{beamed
  region} might be understood as a correspondence between M$_{BH}$ and
Doppler factor, being the latest mainly dependent on the speed of the
jet.  This can be taken as an evidence that \emph{more massive black
  holes produce faster jets} in some BL Lac type objects. Thus BLLacs
outside the \emph{beamed region} must have slower jet speeds.

Figure 9 shows the M$_{BH}$-L$_{5GHz}$ plane where the color-scale
corresponds to the synchrotron peak frequency. A dashed line has been
drawn to approximately separate Doppler boosting dominated BL Lacs,
based on their synchrotron peak frecuency ($\nu_{peak} \approx
10^{16}$ Hz).  As it can be seen, the separation line coincides with
an apparent dichotomy in radio luminosity, most likely influenced by
Doppler boosting effects. This implies that all sources above the line
would shift downwards in Figure 9 if their intrinsic jet luminosity
could be observed. For BL Lacs below the dashed line, Doppler boosting
should not be a major factor and the observed correlation between
black hole mass and luminosity is likely to be intrinsic. If we
consider only sources with $\nu_{peak} > 10^{16}$, we are still
finding that black hole mass correlates  ($\rho =0.3, P = 97 \%$)
with the radio luminosity, roughly of the form M$_{BH} \propto
L_{5GHz} ^{0.5}$. The fact that this slope is shallower than that
found for BL Lacs with $\nu_{peak} < 10^{16}$ (M$_{BH} \propto
L_{5GHz} ^{0.8}$) is merely a consequence of Doppler boosting effects
on the M$_{BH}-luminosity$ relation. This suggests an additional
intrinsic dependence between black hole mass and luminosity for
sources in which luminosity is not Doppler boosting dominated. Whether
host-galaxy properties are somewhat responsible for the intrinsic
luminosity may be gleaned from the stellar-population synthesis that
comes as a byproduct of this study.

\begin{figure}
 \includegraphics[width=\columnwidth]{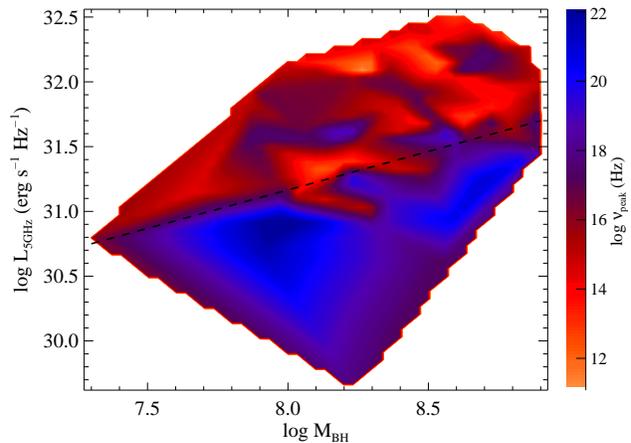}
 \caption{ Black-hole mass - radio luminosity plane, where the color
   scale corresponds to the synchrotron peak frequency. The dashed
   line is drawn at $\nu_{peak} \approx 10^{16}$ Hz, aiming to divide
   Doppler boosting dominated BL Lacs (above the line) from sources
   where Doppler boosting should not be a major factor (below the
   line) in their luminosity. The correlation between
   M$_{BH}$-L$_{5GHz}$ is present for the overall sample of BLLacs.}
 \end{figure}

\section{SUMMARY}

This work presents black hole mass estimations for a sample of 78 BL
Lac type objects drawn from the SDSS-FIRST sample of Plotkin et
al. (2008). The main findings of this work are as follows:
\begin{itemize}

\item Within the sample of BL Lacs considered, HBL and LBL are
  indistinguishable in terms of black hole mass.

\item The black hole mass correlates at high significance level with
  radio luminosity at 5GHz, optical continuum luminosity at 5100 \AA
  ~and X-ray luminosity (see Figure 5).

\item Furthermore, as it can be seen in Figure 6, optical continuum
  and X-ray emission are strongly correlated to the radio luminosity,
  implying that emission in optical and X-ray bands can be associated
  to relativistic jets.

\item When radio luminosity is taken as a third dimension in the plane
  $M_{BH}-\nu_{peak}$, the sample of BL Lacs separates quite clearly
  in regions where luminosity is dominated by intrinsic processes and
  by Doppler beaming effects. The former region is exclusively
  populated by BL Lacs with the heaviest black holes, lowest
  synchrotron peak frequencies and brightest jets.

\item We have found that for a sizeable fraction (40\%) of BLLacs in
  our sample, the emission is dominated by relativistic beaming
  effects. For these sources, the correlation between black hole mass
  and luminosity can be understood as an evidence of a correspondence
  between black hole mass and Doppler boosting factors.

\item If we assume a typical orientation angle for the population of
  BL Lacs, thus the Doppler boosted emission will not strongly
  depend on the viewing angle but mainly on the speed of the
  jet. Therefore, the correlation between the black hole mass and
  Doppler factor suggested by the results found in this work may
  imply that \emph{more massive black holes produce faster jets}.

\item For sources which are not Doppler boosted we still find a
  correlation between black hole mass and radio luminosity, suggesting
  that it must be intrinsic and \emph{more massive black holes produce
    more luminous jets.}

\end{itemize}
 
Hopefully, results presented in this work will provide the impetus for
more observations that look for such correlation among blazars.
\\

We thank the referee for providing useful comments and suggestions
which improved this manuscript.  J. Leon-Tavares acknowledges support
from the Aalto University postdoctoral program offered by the School
of Science and Technology. This work was supported by CONACYT research
grant 54480 (M\'exico). J. Leon-Tavares would like to thank J.P
Torres-Papaqui for his valuable assitance with STARLIGHT.   C.  A\~norve
acknowledges support from the CONACYT program for PhD studies.  The
STARLIGHT project is supported by the Brazilian agencies CNPq, CAPES
and FAPESP and by the France-Brazil CAPES/Cofecub program. Funding for
the SDSS and SDSS-II has been provided by the Alfred P. Sloan
Foundation, the Participating Institutions, the National Science
Foundation, the U.S. Department of Energy, the National Aeronautics
and Space Administration, the Japanese Monbukagakusho, the Max Planck
Society, and the Higher Education Funding Council for England. The
SDSS Web Site is http://www.sdss.org/.

The SDSS is managed by the Astrophysical Research Consortium for the
Participating Institutions. The Participating Institutions are the
American Museum of Natural History, Astrophysical Institute Potsdam,
University of Basel, University of Cambridge, Case Western Reserve
University, University of Chicago, Drexel University, Fermilab, the
Institute for Advanced Study, the Japan Participation Group, Johns
Hopkins University, the Joint Institute for Nuclear Astrophysics, the
Kavli Institute for Particle Astrophysics and Cosmology, the Korean
Scientist Group, the Chinese Academy of Sciences (LAMOST), Los Alamos
National Laboratory, the Max-Planck-Institute for Astronomy (MPIA),
the Max-Planck-Institute for Astrophysics (MPA), New Mexico State
University, Ohio State University, University of Pittsburgh,
University of Portsmouth, Princeton University, the United States
Naval Observatory, and the University of Washington.


\begin{thebibliography}{99}
\bibitem[Abdo et al. (2010)]{abdo2010} Abdo, A. A. et al. 2010, ApJ submitted, (arXiv:1002.0150)
\bibitem[Arshakian et al. (2005)]{arshakian05} Arshakian,T. G., Chavushyan, V. H., Ros, E., Kadler M. \& Zensus, J. A. 2005, MmSAI, 76, 35
\bibitem[Arshakian et al. (2010)]{arshakian10} Arshakian, T. G., Le\'on-Tavares, J., Lobanov, A. P., Chavushyan, V. H., Shapovalova, A. I., Burenkov, A. N. \& Zensus, J. A. 2010, MNRAS, 401, 1231
\bibitem[ Asari et al. (2007)]{asari07} Asari, N. V., Cid Fernandes, R., Stasińska, G., Torres-Papaqui, J. P., Mateus, A., Sodr\'e, L., Schoenell, W., Gomes, J. M. . 2007, MNRAS, 381, 263
\bibitem[Blandford \& Konigl (1979)]{blandford79} Blandford, R. D., \& Konigl, A. 1979, ApJ, 232, 34
\bibitem[Bertin \& Arnouts (1996)]{bertin96} Bertin, E \& Arnouts, S. 1996, A\&AS, 117, 393
\bibitem[Bruzual, G \& Charlot (2003)]{bc03} Bruzual, G \& Charlot, M. J. 2003, MNRAS, 344, 1000 
\bibitem[Cardelli et al. (1989)]{cardelli89}Cardelli J. A., Clayton, G. C. \&  Mathis, J. S.  1989, ApJ, 345, 245
\bibitem[Cid Fernandes et al.(2004)]{2004MNRAS.355..273C} Cid Fernandes, R., Gu, Q., Melnick, J., Terlevich, E., Terlevich, R., Kunth, D., Rodrigues Lacerda, R., \& Joguet, B.\ 2004, MNRAS, 355, 273
\bibitem[Cid Fernandes et al.(2005)]{2005MNRAS.358..363C} Cid Fernandes, R., Mateus, A., Sodr{\'e}, L., Stasi{\'n}ska, G., \& Gomes, J.~M.\ 2005, MNRAS, 358, 363
\bibitem[Cid Fernandes et al.(2007)]{2007MNRAS.375L..16C} Cid Fernandes, R., Asari, N.~V., Sodr{\'e}, L., Stasi{\'n}ska, G., Mateus, A., Torres-Papaqui, J.~P., \& Schoenell, W.\ 2007, MNRAS, 375, L16
\bibitem[Dunlop et al. (2003)]{dunlop03} Dunlop, J. S., McLure, R. J., Kukula, M. J., Baum, S. A., O'Dea, C. P. \&  Hughes, D. H. 2003, MNRAS, 340, 1095 
\bibitem[Falomo et al. (2000)]{falomo00} Falomo, R., Scarpa, R., Treves, A. \&  Urry, C. M. 2000, ApJ, 542, 731
\bibitem[Falomo et al. (2002)]{falomo02} Falomo, R., Kotilainen, J. K., Treves, A. 2002, ApJL,596, 35

\bibitem[Falomo et al. (2003)]{falomo03} Falomo, R., Kotilainen, J. K., Carangelo, N. \&  Treves, A. 2003, ApJ, 595, 624
\bibitem[Ferrarese  et al. (2001)]{ferrarese01} Ferrarese, L., Pogge, R. W., Peterson, B. M., Merritt, D., Wandel, A., Joseph, C. L. 2001, ApJL, 555, 79
\bibitem[ Ili\'c et al. (2008)]{ilic08} Ili\'c, D., Popovi\'c, L. \v C., Le\'on-Tavares, J., Lobanov, A. P., Shapovalova, A. I., Chavushyan, V. H. 2008, MmSAI, 79,1105 
\bibitem[ Jester et al. (2005)]{jester05} Jester, S. et al. 2005 ,AJ, 130, 873
\bibitem[Kaspi  et al. (2000)]{kaspi00}	Kaspi, S., Smith, P. S., Netzer, H., Maoz, D., Jannuzi, B. T. \&  Giveon, U. 2000 ApJ, 533, 631

\bibitem[Komatsu et al. (2009)] {komatsu09} Komatsu, E., et al. 2009, ApJS, 180, 330

\bibitem[Le\'on-Tavares et al. (2010)] {lt09} Le\'on-Tavares, J.,
  Lobanov, A. P., Chavushyan, V. H., Arshakian, T. G., Doroshenko,
  V. T. et al. 2010, ApJ, 715, 355

\bibitem[ Lister et al. (2009)]{lister09} Lister, M. L., Cohen, M. H., Homan, D. C., Kadler, M., Kellermann, K. I., Kovalev, Y. Y., Ros, E., Savolainen, T. \& Zensus, J. A. 2009, ApJL, 696, 22
\bibitem[ McLure  \&  Dunlop  (2002)]{mclure02} McLure, R. J. \&  Dunlop, J. S. , MNRAS, 331, 795 
\bibitem[Mateus et al. (2006)]{mateus06} Mateus, A., Sodre, L., Cid Fernandes, R., Stasinska, G., Schoenell, W., \&  Gomes, J. M. 2006, MNRAS, 370, 721
\bibitem[Nieppola et al. (2006)]{nieppola06} Nieppola, E., Tornikoski, M. \& Valtaoja, E. 2006, A\&A,445, 441 
\bibitem[Nieppola et al.  (2008)]{nieppola08} Nieppola, E., Valtaoja, E., Tornikoski, M., Hovatta, T. \& Kotiranta, M. 2008 , A\&A, 488, 867 

\bibitem[Nilsson et al. (2003)]{nilsson}Nilsson, K., Pursimo, T., Heidt, J., Takalo, L. O., Sillanp\"a\"a, A., \& Brinkmann, W. 2003 A\&A, 400, 95

\bibitem[O'Dowd \& Urry (2005)]{odowd} O'Dowd, M. \&  Urry, C. M. 2005, ApJ, 627, 97

\bibitem[Padovani \& Giommi (1995)]{padovani95} Padovani, P., Giommi, P. 1995, ApJ, 444, 567 
\bibitem[Peng  et al. (2002)]{peng02}Peng, C. Y., Ho, L. C., Impey, C. D. \&  Rix, H. 2002, AJ, 124, 266
\bibitem[Perlman et al. (1996)]{perlman96} Perlman, E. S., Stocke, J. T., Wang, Q. D. \& Morris, S. L., 1996, ApJ, 456, 451

\bibitem[Peterson et al. (2004)]{peterson} Peterson, B. M., Ferrarese, L., Gilbert, K. M., Kaspi, S., Malkan, M. A., et al. 2004, ApJ, 613, 682

\bibitem[Plotkin  et al. (2008)]{plotkin08} Plotkin, R. M., Anderson, S. F., Hall, P. B., Margon, B., Voges, W., Schneider, D. P., Stinson, G., York, D. G. 2008, AJ, 135, 2453 
\bibitem[Poggianti (1997)]{poggianti97} Poggianti, B. 1997 A\&AS, 122, 399


\bibitem[Savolainen et al. (2010)]{savolainen10} Savolainen, T., Homan, D. C., Hovatta, T., Kadler, M., Kovalev, Y. Y.; Lister et al. 2010, A\& A, 512, 24

\bibitem[Scarpa et al. (2000)]{scarpa} Scarpa, R., Urry, C. M.,
  Padovani, P., Calzetti, D. \& O\'Dowd, M.  2000, ApJ, 532, 740


\bibitem[Schlegel et al. (1998)]{schlegel98} Schlegel, D.J., Finkbeiner D.P., \& Davis M. 1998, ApJ, 500, 525
\bibitem[Tremaine et al. (2002)]{tremanine02} Tremaine, S. et al. 2002, AJ, 574, 740
	

\bibitem[Tornikoski et al. (2010)]{tornikoski10}Tornikoski, M., Nieppola, E., Valtaoja, E., Le\'on-Tavares, J. \& L\"ahteenm\"aki, A. 2010, Proceedings of the Workshop "Fermi meets Jansky - AGN in Radio and
Gamma-Rays", Savolainen, T., Ros, E., Porcas, R.W. \& Zensus, J.A.
(eds.), MPIfR, Bonn, June 21-23 2010


\bibitem[Torrealba et al. (2008)]{torrealba08} Torrealba, J.,
  Chavushyan, V. H., Arshakian, T. G., Cruz-Gonz\'alez, I., Ros, E.,
  Zensus, J. A., Bertone, E., Rosa-Gonz\'alez, D. 2008, The Nuclear
  Region, Host Galaxy and Environment of Active Galaxies (Eds. Erika
  Ben\'{\i}tez, Irene Cruz-Gonz\'alez, \& Yair Krongold) Revista
  Mexicana de Astronom\'{\i} a y Astrof\'{\i}sica (Serie de Conferencias)
  Vol. 32, pp. 48-48


\bibitem[Urry et al. (2000)]{urry} Urry, C. M., Scarpa, R., O'Dowd, M., Falomo, R., Pesce, J. E., Treves, A. 2000, ApJ, 532, 816

\bibitem[Valtaoja  et al. (2008)]{valtaoja08} Valtaoja, E., Lindfors, E., Saloranta, P.-M., Hovatta, T., L\"ahteenm\"aki, A., Nieppola, E., Torniainen, I. \& Tornikoski, M. 2008, ASPC, 386, 388


\bibitem[Vestergaard (2009)]{vestergaard09}Vestergaard M., 2009,
  Spring Symposium on “Black Holes” (Cambridge University Press) in
  press, (astro–ph/0904.2615)

\bibitem[Woo et al. (2005)]{woo05} Woo, J.-H., Urry, C. M., Van der Marel, R. P., Lira, P., \& Maza, J. 2005, ApJ, 631, 762


\end{thebibliography}
\end{document}